\documentclass[pra,twocolumn,showpacs,preprintnumbers,amsmath,amssymb,superscriptaddress]{revtex4}

\usepackage[usenames]{color}
\usepackage{graphicx}
\usepackage{epsfig}
\usepackage{amsmath}

\begin{document}

\title{Dissipation control in cavity QED with oscillating mode-structures}
%
\author{I. E. Linington}
\affiliation{Department of Physics and Astronomy, University of Sussex,
  Falmer, Brighton, BN1 9QH, United Kingdom}
\affiliation{Department of Physics, Sofia University, James Bourchier 5 blvd, 1164 Sofia,
Bulgaria}
\author{B.M. Garraway} 
\affiliation{Department of Physics and Astronomy, University of Sussex,
  Falmer, Brighton, BN1 9QH, United Kingdom}

\begin{abstract}
We demonstrate how a
  time-dependent dissipative environment may be used as a tool for controlling the quantum state of a two-level atom.
  In our model system the frequency and coupling strength associated with microscopic reservoir modes are modulated, while the principal features of the reservoir structure remain fixed in time. Physically, this may be achieved by containing a static atom-cavity system inside an oscillating external bath. 
We show that it is possible to dynamically decouple the atom from its environment, despite the fact that the two remain resonant at all times. This can lead to Markovian dynamics, even for a strong atom-bath coupling, as the atomic decay becomes inhibited into all but a few channels; the reservoir occupation spectrum consequently acquires a sideband structure, with peaks separated by the frequency of the environmental modulation. The reduction in the rate of spontaneous emission using this approach can be significantly greater than could be achieved with an oscillatory atom-bath detuning using the same parameters.
\end{abstract}
\date{\today}

\maketitle

\section{Introduction}
\label{introduction}
%
It is well known that the ultimate origin of dissipation is the unavoidable interaction between a system and its surroundings. For quantum systems, a distinction is usually made between \emph{decay}, during which energy leaks into the environment, and \emph{decoherence}, whereby correlations established with the surroundings rapidly wash-out any coherence between certain elements of the system \cite{zurek2003}. Decoherence generally takes place on a much shorter timescale than decay, and this timescale is governed by the separation in phase-space of different components of the system density operator. However, for a single two-level atom, the timescale governing both decay and decoherence is related to the rate of spontaneous emission, which simplifies matters and thus makes this an ideal quantum system in which to study dissipative effects.

The idea of using the surroundings to control atomic decay has been pursued ever since the first studies using nearby conducting surfaces and cavities \cite{purcell1946,kleppner1981,goy1983,jhe1986} and experimental control has now reached the point where an atom can be strongly-coupled to a cavity in which it is placed. This technological development has allowed for many useful applications of the atom-cavity interaction, including the generation of entanglement \cite{rauschenbeutel2000}, creation of number states of the electromagnetic field \cite{weidinger1999,varcoe2000} and photons on demand \cite{keller2004,lounis2005}, observation of quantum jumps of the electromagnetic field \cite{gleyzes2007} and experimental tests of non-locality \cite{milman2005} among many others \cite{walther2006}. Nevertheless, interaction with the environment is traditionally seen as a negative feature since this is the root-cause of dissipation and there is currently a huge theoretical and experimental effort aimed at controlling dissipative effects in cavity quantum optics.

One way to achieve this goal is to limit the extent of the system-environment coupling, although ultimately this will always still prevail at some level. Another promising approach is to exploit certain symmetries in the system-environment coupling, so as to confine the dynamics to a (decoherence-free) subspace of the overall Hilbert space in which dissipative effects cancel out \cite{zanardi1997,beige2000}.
A third method, and the approach we shall pursue here, is to dynamically modify the system's surroundings so as to control the \emph{effects} of the dissipative couplings rather than reducing their magnitude \cite{viola1999,agarwal2001,kofman2001, pechen2006, pechen2008}. The use of dynamically engineered reservoirs to control dissipation is appealing, since this technique is non-invasive and furthermore does not require individual components of the system to be addressed separately.

To the lowest order of approximation, an atom's environment consists simply of the electromagnetic field modes into which it can emit. We should therefore expect that changing some property of \emph{all} of these modes should affect the process of atomic decay, and these effects are well documented \cite{law1995, agarwal1999, kofman2001}. It is less clear what will happen in the more subtle case where the properties of individual modes are altered, but in a controlled way, such that, for example, the combined effect is to keep all macroscopic reservoir structure static in time. It is tempting to think that this type of manipulation might only produce observable effects in strongly-coupled atom-cavity systems, since it is only the memory kernel for the atom-bath interaction which changes and is dynamically modified under this model. However, it has recently been shown (in Ref. \cite{linington2006} for a linear increase of all reservoir frequencies) that this form of reservoir manipulation can be used to control the rate of spontaneous emission for both strong and also weak atom-reservoir coupling.

In the current paper we investigate the case of an oscillatory manipulation of all reservoir mode frequencies and coupling strengths. This can be achieved by modifying the length of a cavity, and hence the mode frequencies, in an oscillatory way, (further details are given in section \ref{observable_effects} of this article). From a practical perspective, an oscillating mode-structure has several advantages over a linear chirp of the reservoir frequencies: (i) piezoelectric actuators used to modify the cavity length typically perform favourably in an oscillatory regime; (ii) the required changes in the cavity length are periodic and do not grow large at long times, unlike in the linear-chirp case.

The remainder of the article is arranged as follows. In section \ref{general_dynamic_environment}, the mathematical model is introduced and some general observations are made regarding the applicability of dynamic environments to the control of decoherence and decay. In section \ref{dynamic_res_model}, the detailed form of the microscopic reservoir structure is given for an \emph{oscillatory modulation} of the reservoir mode frequencies  and closed-form expressions for the atomic decay rate and emission spectrum are derived. These results are tested against numerical  simulations in section \ref{sinusoidal_example} for the specific case of a sinusoidal modulation of the bath mode frequencies. A particular physical realisation of the model and the potential size of observable effects are examined in detail in section \ref{observable_effects}. It is found that our scheme can give rise to a surprisingly large reduction in the rate of dissipation and that the method studied here compares favourably with other, more basic reservoir manipulations studied previously (e.g. \cite{law1995, janowicz2000, agarwal1999, blohe1996}). Finally, in section \ref{conclusions} we conclude our findings.
%
\section{General dynamic environments for a two-level atom}
\label{general_dynamic_environment}
%
We consider a two-level atom with transition frequency
$\omega_{0}$ and lower and upper states
$\vert{}0\rangle$ and $\vert{}1\rangle$, coupled to a zero-temperature bath of electromagnetic field modes, which together  constitute a reservoir for the atomic
decay. The reservoir is engineered so that the individual bath mode frequencies
$\omega_{k}(t)$ and also the coupling $g_{k}(t)$ between the atomic transition 
$\vert{}0\rangle\leftrightarrow\vert{}1\rangle$ and the $k^{th}$ mode of the radiation field are \emph{time-dependent}. We initially consider a discrete bath. Without loss of generality, all couplings are chosen to be real, since any time-dependent phase in the coupling $g_{k}(t)$ can be transferred onto the time-dependence of the corresponding mode frequency $\omega_{k}(t)$ and as usual, constant phase terms can be absorbed into the basis states.
The Hamiltonian for the composite atom-reservoir system in the rotating-wave approximation (with $\hbar=1$) is:
\begin{align}
\hat{H}(t) = \;&
\omega_{0}\hat{\sigma}^{+}\hat{\sigma}^{-}+\sum_{k}\omega_k(t)\left(\hat{b}_{k}^{\dagger}\hat{b}_{k}+1/2\right)
 \nonumber  \\
& 
+\sum_{k}g_k(t)\left(\hat{\sigma}^{-}\hat{b}_{k}^{\dagger}+\hat{b}_{k}\hat{\sigma}^{+}\right),
\label{Hamiltonian}
\end{align}
with raising and lowering operators
\mbox{$\hat{\sigma}^{+}=\vert{}1\rangle\langle{}0\vert$} and \mbox{$\hat{\sigma}^{-}=\vert{}0\rangle\langle{}1\vert$}.
Due to the oscillation of the cavity, individual mode-frequencies may pass through resonance on a timescale which is very rapid from the point-of-view of the atom and may even be much shorter than the bath correlation time. However, we note that in any practical setting, the rate of change of reservoir mode frequencies is never fast enough to create photons in the reservoir by itself (in the sense of refs.
\cite{janowicz1998,dodonov1996}). Thus, the cavity field will adiabatically
follow the motion of its end-mirror, and so the bath creation and annihilation operators, 
$\hat{b}_{k}^{\dagger}$ and $\hat{b}_{k}$, depend only on the index $k$ (not explicitly on time).

It is convenient to move to an interaction picture, in which the effect of the first two terms in equation (\ref{Hamiltonian}) are factored into the energy basis states. In this rotated basis, the Hamiltonian takes the form
\begin{align}
\hat{H}_{I}(t) = &
\sum_{k}g_{k}(t)\bigg[\hat{\sigma}^{-}\hat{b}_{k}^{\dagger}\exp\left(i\int_{0}^{t}
\big[\omega_k(\tau) - \omega_{0}\big]
d\tau\right)  
\nonumber \\
& +\hat{\sigma}^{+}\hat{b}_{k}\exp\left(-i\int_{0}^{t}\big[\omega_k(\tau') - \omega_{0}\big]d\tau'\right)\bigg].
\label{H_int_osc}
\end{align}
Since the Hamiltonian commutes with $\hat{N}=\hat{\sigma}^{+}\hat{\sigma}^{-}+\hat{b}_{k}^{\dagger}\hat{b}_{k}$, the total number of energy quanta is a constant of the motion. We choose the atom to be in its excited state at $t=0$ and thus the state-vector at any later time is simply a superposition of those energy eigenstates corresponding to a single energy quantum:
\begin{align}
\left\vert\psi_{I}(t)\right\rangle = & 
c_{a}(t)\left\vert1\right\rangle\otimes\left\vert\ldots0\ldots\right\rangle \label{state_vector}
\nonumber \\ & +
\sum_{k}c_{k}(t)\left\vert0\right\rangle\otimes\left\vert\ldots1_{k}\ldots\right\rangle.
\end{align}
Using the Schr\"odinger equation, the equation of motion of the state-vector coefficients is found to be
\begin{align}
i\frac{\partial{}c_{a}(t)}{\partial{}t} = & \sum_{k}g_{k}\left(t\right)\exp\left(-i\int_{0}^{t}\big[\omega_k(\tau) - \omega_{0}\big]d\tau\right)c_{k}(t)
\label{chirp_dynamics1_osc} 
\\
i\frac{\partial{}c_{k}(t)}{\partial{}t} = & g_{k}\left(t\right)\exp\left(i\int_{0}^{t}\big[\omega_k(\tau) - \omega_{0}\big]d\tau\right)c_{a}(t).
\label{chirp_dynamics2_osc}
\end{align}
Since we are interested in controlling the atomic state, it is convenient to eliminate the environmental variables, $c_{k}(t)$, from equations (\ref{chirp_dynamics1_osc}) and (\ref{chirp_dynamics2_osc}) giving the following integro-differential equation for the atomic state:
\begin{align}
\frac{\partial{}c_{a}(t)}{\partial{}t} = & -\int_{0}^{t}K(t,t')c_{a}(t')\:dt',
\label{general_atomic_dynamics}
\end{align}
with
\begin{align}
K(t,t') = & \sum_{k}g_{k}(t')g_{k}(t)\exp\left(-i\int_{t'}^{t}\big[\omega_k(\tau) - \omega_{0}\big]d\tau\right).
\label{kernel_osc}
\end{align}
We note that the atomic dynamics are determined solely by the behaviour of the kernel, $K(t,t')$, which in turn is dependent only on the properties of the atom's surroundings:
The principal idea underlying the use of dynamic environments is that by manually altering the reservoir properties, $g_{k}(t)$ and $\omega_{k}(t)$, it is possible to control the evolution of the atomic state. In other words, the inevitable coupling between system and surroundings can be used to our advantage, since changes in the reservoir have an effect on the atom, and this can be used to shape the atomic state.

As a preliminary observation, we note that imposing a dynamic structure on the reservoir can only give rise to observable effects if the atom-field coupling is \emph{structured}, by which we mean that the product $\rho_{k}g_{k}^{2}$, varies significantly with frequency, where $\rho_{k}$ is the reservoir density of states. To see this, we note that in the converse situation of a completely structureless reservoir, with $\rho_{k}g_{k}(t)^{2}=\rho_{0}g_{0}^{2}$ (constant), then whatever the modulation function $f(t)$ we choose for the mode frequencies \footnote{We assume that all the modes are modulated identically, so as to keep the density of states unchanged.}
\begin{align}
\omega_{k}(t) = & \omega_{k}(0)+f(t),
\end{align}
the kernel, (\ref{kernel_osc}), reduces to
%
\begin{align}
K(t,t') = &  g_{0}^{2}\sum_{k}\exp\left(-i\int_{t'}^{t}\big[\omega_{k}(\tau) - \omega_{0}\big]d\tau\right)
\nonumber \\
= &  g_{0}^{2}\left[\sum_{k}\exp\Big(-i\left(\omega_{k}(0) - \omega_{0}\right)\left(t-t'\right)\Big)\right]
\nonumber \\
& \hspace{5mm}\times\exp\left(-i\int_{t'}^{t}f(\tau)d\tau\right)
\nonumber \\
\approx & 2\pi\rho_{0}g_{0}^{2}\delta(t-t').
\end{align}
%
The population of the atomic excited state thus decays exponentially at exactly the rate predicted by Fermi's golden rule for a static reservoir:
\begin{align}
\frac{\partial{}c_{a}(t)}{\partial{}t} = & -\pi\rho_{0}g_{0}^{2}c_{a}(t).
\label{FGR_result}
\end{align}
Equation (\ref{FGR_result}) holds, regardless of the frequency modulation $f(t)$ that is imposed on the reservoir modes. Therefore, even though the atom is resonant with different field-modes at different times during its evolution, the overall effect of the reservoir is unchanged by modulating the mode frequencies alone. Dynamically modulating the microscopic properties of a reservoir can only have an observable effect if the system-reservoir coupling is a structured function of frequency. However, we note that the requirement of a structured reservoir is not synonymous with strong coupling between the system and its environment, since the distinction between strong coupling, reservoir structure and non-Markovian dynamics is more subtle  for dynamic reservoirs than the usual static case. (For further details see \cite{linington2006}.)

\section{Dynamic Reservoir Model}
\label{dynamic_res_model}
\begin{figure}[!h]
\includegraphics[width=0.75\columnwidth]{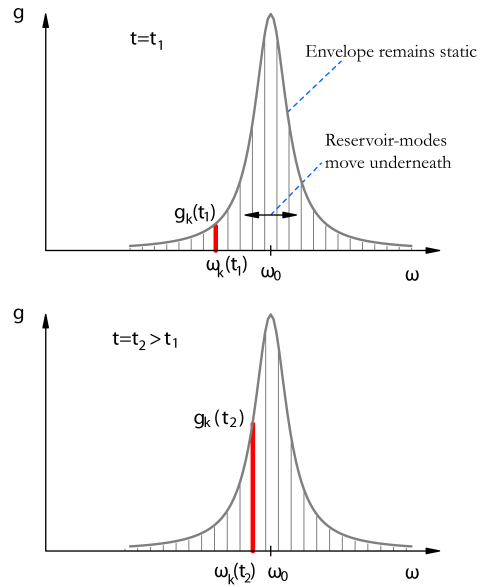}
\caption{
 This figure corresponds to the model studied in section
 \ref{dynamic_res_model} and shows reservoir mode couplings $g_k(t)$ (see
 equation (\ref{H_int_osc})) at two different times $t_1$ and $t_2$. The
 envelope of the couplings remains the same as the reservoir mode frequencies
 vary in time.  Thus the coupling of the atom to an individual mode changes in
 time.  The thick line indicates the same mode $k$ at two different times,
 $t_{1}$ and $t_{2}$. Since at time $t_{2}$ this mode has moved to a
 different frequency under the static envelope, the coupling has changed
 accordingly.}
\label{modesFig_osc} 
\end{figure}

%
Having established that the use of dynamic environments requires the system-reservoir coupling to be spectrally dependent, we now turn to the specific form that this dependence will take. 
In order to isolate the novel effects associated with dynamically structured reservoirs, we choose to investigate a model for which the \emph{macroscopic} properties of the reservoir remain fixed. (See figure \ref{modesFig_osc}.) This means that the individual coupling strengths, $g_{k}(t)$ must vary in time, to match the local reservoir structure at the changing frequency, $\omega_{k}(t)$. In this way the envelope of the coupling strengths can be kept static. Put another way, the microscopic atom-mode couplings conspire in such a manner that the reservoir structure function
$\rho_{k}\vert{}g_{k}(t)\vert^{2}$ must be expressible as a pure function of $\omega_{k}(t)$ only, with no explicit dependence on the time $t$. 
%
%
For a simple cavity model, we assume that the single time reservoir structure function $\rho_{k}\vert{}g_{k}(t)\vert^{2}$ is a Lorentzian, with width $\gamma$, centred on the atomic transition frequency $\omega_{0}$,
i.e.
\begin{align}
\rho_{k}\vert{}g_{k}(t)\vert^2 = & 
\frac{D^2\gamma/\pi}{\gamma^2+(\omega_{k}(t) - \omega_{0})^2}.
\label{rho_g_squared}
\end{align}
For a static bath, where
$\rho_{k}\vert{}g_{k}(t)\vert^2$ is not a function of time, the
Lorentzian (\ref{rho_g_squared}) is a common choice for the reservoir
structure \cite{barnett1997,lang1973} and the resulting dynamics have been
well explored (see, for example, \cite{garraway1997, lambropoulos2000}). In the weak coupling limit, the
Lorentzian reservoir structure ensures exponential decay of atomic population.

Equation (\ref{rho_g_squared}) describes both the time-dependent coupling of a single bath mode, $\omega_{k}(t)$, and also the instantaneous coupling between the atom and the whole bath of modes at time $t$. Since this structure has no explicit dependence on $t$ (i.e. only the implicit time-dependence contained in $\omega_{k}(t)$) the envelope of these couplings is fixed in time; as in the static case, the atom always has some resonant modes with which it can exchange a photon but these are now different bath modes at different times.
We note that the weight, $D$, of the Lorentzian, which is defined through the relation
\begin{equation}
\sum_{k}\vert{}g_{k}(t)\vert^{2} = D^{2},
\end{equation}
also remains static in time with this choice of couplings.

Of course, the kernel, (\ref{kernel_osc}) does not feature
$\rho_{k}\vert{}g_{k}(t)\vert^{2}$, but the two-time product
$\rho_{k}g_{k}(t)g_{k}(t')$. However, as already mentioned,
$g_{k}(t)$ may be chosen to be real. Therefore the two-time product follows
immediately as
%
\begin{align}
\rho_{k}g_{k}(t)g_{k}(t') = & \frac{D^2\gamma}{\pi\sqrt{\left(\gamma^2+(\omega_{k}(t) - \omega_{0})^2\right)}}
\nonumber \\
&\times\frac{1}{\sqrt{\left(\gamma^2+(\omega_{k}(t') - \omega_{0})^2\right)}}.
\label{rho_g(t)_g(t')_osc}
\end{align}
%
To proceed with the analysis, a specific form for the time
dependence of the modes must also be chosen. We assume that all the reservoir modes experience an identical frequency-modulation, which is periodic in time, with period $\Omega$
\begin{equation}
\omega_{k}(t) = \omega_{k}(0) + f(t).
\label{frequency_definition_osc}
\end{equation}
As well as the modulation frequency, $\Omega$, it will also prove helpful to introduce the \emph{modulation depth}
\begin{align}
d = & \frac{1}{2}\left(f_{\textrm{max}}-f_{\textrm{min}}\right),
\end{align}
in order to characterise the amplitude of the frequency manipulation. We also note that the specific form $f(t)=d\sin(\Omega{}t)$ will be considered in detail in section \ref{sinusoidal_example}.

By choosing $f(t)$ to be periodic, we ensure that the dynamically acquired phase-term  arising from the reservoir manipulation is also periodic, and thus has a discrete Fourier-decomposition:
%
\begin{align}
\exp\left(-i\int_{0}^{t}f(\tau)d\tau\right) = & \sum_{n=-\infty}^{\infty}F_{n}e^{-in\Omega{}t}
\end{align}
with
\begin{align}
F_{n} = & \frac{\Omega}{2\pi}\int_{0}^{\frac{2\pi}{\Omega}}\exp\left(-i\int_{0}^{t}f(\tau)d\tau\right)e^{-in\Omega{}t}\;dt.
\label{Fourier_decomposition}
\end{align}
%
For this specific frequency manipulation we note that the Hamiltonian (\ref{H_int_osc}) may be re-expressed as follows:
\begin{widetext}
\begin{align}
\hat{H}_{I}(t) = &
\sum_{k}g_{k}(t)\bigg[\hat{\sigma}^{-}\hat{b}_{k}^{\dagger}\exp\left(i\left(\omega_k(0)
- \omega_{0}\right)t + i\int_{0}^{t}f(\tau)d\tau\right) 
 \nonumber \\
&  \hspace{14mm}+\hat{\sigma}^{+}\hat{b}_{k}\exp\left(-i\left(\omega_k(0)
- \omega_{0}\right)t - i\int_{0}^{t}f(\tau')d\tau'\right)\bigg]
\nonumber \\
= & \sum_{n=-\infty}^{n=\infty}\Bigg\{\sum_{k}F_{n}g_{k}(t)\hat{\sigma}^{-}\hat{b}_{k}^{\dagger}e^{i\left(\omega_k(0)
- \omega_{0} - n\Omega\right)t}
+F_{n}^{*}g_{k}(t)\hat{\sigma}^{+}\hat{b}_{k}e^{-i\left(\omega_k(0)
- \omega_{0} - n\Omega\right)t}\Bigg\}.
\label{H_int_osc_interpreted}
\end{align}
\end{widetext}
Written in this way, we see that the chosen dynamic reservoir structure (figure \ref{modesFig_osc}) is effectively the same as an infinite collection of independent reservoirs, indexed by $n$, which have static mode-frequencies, and time-dependent couplings, $F_{n}g_{k}(t)$. (These are the terms in curly braces in equation (\ref{H_int_osc_interpreted}).) This picture is illustrated in figure \ref{detuned_moving_baths}.
The $n^{th}$ sideband peak in the reservoir structure has central frequency $\omega_{0}+n\Omega$. From the definition of the coupling strengths, (\ref{rho_g_squared}), we see that the $n^{th}$ sideband has the largest coupling when 
\begin{align}
f(t) + n\Omega \approx &  0.
\label{dynamic_resonance_condition}
\end{align}
Since the condition, (\ref{dynamic_resonance_condition}) is different for each sideband, the atom decays into different sidebands at different times. The total decay-rate is therefore determined by the decay-rates into each individual reservoir, as shown in the following section.

\subsection{Atomic decay rates}
\label{atomic_decay_rates}
%
Using equations (\ref{rho_g(t)_g(t')_osc}) and (\ref{frequency_definition_osc}), we can now write an explicit expression for the kernel
\begin{widetext}
\begin{align}
K(t,t') = &
\frac{D^2\gamma}{\pi\rho_{k}}\sum_{k}
\frac{\exp\left[-i\left(\omega_{k}(0) -
      \omega_{0}\right)\left(t-t'\right)\right]}{\sqrt{\left(\gamma^2+\left(\omega_{k}(0) - \omega_{0}+f(t)\right)^{2}\right)\left(\gamma^2+\left(\omega_{k}(0) -
     \omega_{0}+f(t')\right)^2\right)}}
\cdot{}\exp\left(-i\int_{t'}^{t}f(\tau)d\tau\right).
\label{ca_integro_diff_osc}
\end{align}
It is convenient now to move to a continuum limit for the mode frequencies. For economy, we choose to write the initial frequencies simply as $\omega_{k}(0)\equiv\omega$, which gives:
\begin{align}
K(t,t') = &
\frac{D^2\gamma}{\pi}
\int_{-\infty}^{\infty}\frac{\exp\left[-i\left(\omega -
      \omega_{0}\right)\left(t-t'\right)\right]}{\sqrt{\left(\gamma^2+\left(\omega - \omega_{0}+f(t)\right)^{2}\right)\left(\gamma^2+\left(\omega -
     \omega_{0}+f(t')\right)^2\right)}}d\omega\cdot{}\exp\left(-i\int_{t'}^{t}f(\tau)d\tau\right).
\label{Kernel_osc_continuous}
\end{align}
\end{widetext}

\begin{figure}[!h]
\includegraphics[width=0.9\columnwidth]{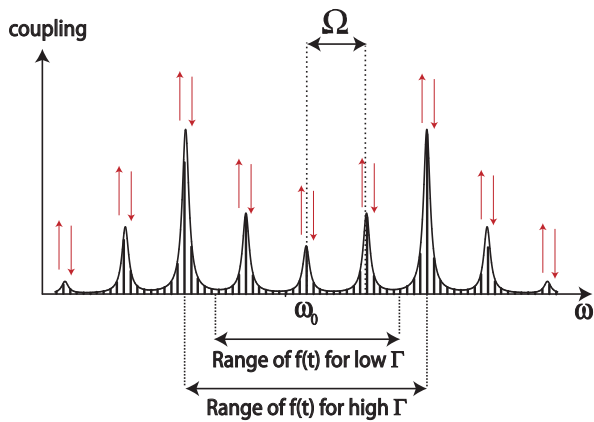}
\caption{By re-writing the Hamiltonian (\ref{H_int_osc}) in the form
(\ref{H_int_osc_interpreted}), the model studied in this paper may be interpreted as describing the decay into a
collection of reservoir sidebands, evenly spaced by angular frequency $\Omega$. In this
picture, all reservoir mode frequencies are static in time, but the coupling
profile of each bath is modulated according to
(\ref{rho_g(t)_g(t')_osc}) (vertical arrows). Each bath passes through resonance at
a different time, specified by equation (\ref{dynamic_resonance_condition}).
As discussed in below equations (\ref{Gamma_n_osc}-\ref{sideband_couplings}), the atomic decay-rate, $\Gamma_{\infty}$, can be increased or decreased by tuning the extremal values of $f(t)$ towards or away from the sideband frequencies, $\vert{}f(t)\vert\approx{}n\Omega$.}
\label{detuned_moving_baths} 
\end{figure}

The kernel contains information about the `memory' of the atomic dynamics, in the sense of equation (\ref{general_atomic_dynamics}). As a special case, we note that when the kernel decays sharply away from $t'\sim{}t$, memory effects are absent and the Markov approximation can be applied, to give:
\begin{align}
\frac{\partial{}c_{a}(t)}{\partial{}t} \approx & -\frac{\Gamma_{\infty}}{2}c_{a}(t),
\end{align}
with
\begin{align}
\Gamma_{\infty} = & \lim_{t\rightarrow\infty}\left\{\frac{2}{t}\Re\int_{0}^{t}dt_{1}\int_{0}^{t_{1}}dt_{2}\;\;\;K(t_{1},t_{2})\right\}.
\label{Gamma_Kernel}
\end{align}
For the Markov approximation to hold well, the Kernel must decay on a time-scale which is very short compared to the typical time-scale of atomic dynamics. For static reservoirs this requires that the system-reservoir coupling is weak (i.e. $D\ll\gamma$). However, for dynamic reservoirs the atomic behaviour can be Markovian, even when the atom-reservoir coupling is strong, as long as the two-time product in the denominator of $K(t_{1},t_{2})$ decays sufficiently fast. Thus, for dynamically manipulated environments, the atomic behaviour can be Markovian even in the strong-coupling regime \footnote{We note that the very short time behaviour remains non-exponential, as in the quantum Zeno effect. However, the extent of this behaviour is so short in time that it does not disturb the results in the regime we discuss here.}. The results that follow in this paper therefore hold when \emph{either} of the two below conditions are satisfied:
\begin{enumerate}
\item
Weak-coupling:$\;\;\;\;D\ll\gamma$
\item
Fast rate of change of reservoir mode-frequencies:$\;\;\;\;d\Omega\ll\gamma^{2},D^{2}$.
\end{enumerate}

Our task is now to solve for the atomic decay-rate, $\Gamma_{\infty}$, and to see how this can be controlled by altering the properties of the modulation-function $f(t)$ which characterises the environmental manipulation. To this end, we note that (\ref{Gamma_Kernel}) can be written in the compact form:
\begin{align}
\Gamma_{\infty} = & \lim_{t\rightarrow\infty}\Bigg\{\frac{D^{2}\gamma}{\pi{}t}\int_{-\infty}^{\infty}d\omega\left(\int_{0}^{t}dt_{1}\;P(\omega,t_{1})\right)
\nonumber \\
& \hspace{18mm}\times\left(\int_{0}^{t}dt_{2}\;P(\omega,t_{2})\right)^{*}\Bigg\},
\label{rewrite_Gamma_varying_gs}
\end{align}
with
\begin{align}
P(\omega,t)  = & \left\{\frac{\exp\left(- i\int_{0}^{t}f(\tau)\;d\tau\right)}{\sqrt{\gamma^2+\left(\omega_{} - \omega_{0}+f(t)\right)^{2}}}\right\}
\nonumber\\
&\times\exp\left[-i\left(\omega-\omega_{0}\right)t\right].
\label{P_def_osc}
\end{align}
The mathematical steps involved in proceeding from equation (\ref{Gamma_Kernel}) to equation (\ref{rewrite_Gamma_varying_gs}) include a change of limits, a re-assignation of dummy indices and use of the fact that $\Gamma_{\infty}$ is defined to be real.
The term $\int_{0}^{t_{1}}P(\omega,t_{1})\;dt_{1}$ in equation (\ref{rewrite_Gamma_varying_gs}) is thus only significant if the two oscillating factors which make up $P(\omega,t)$ interfere constructively over many cycles. 
Due to the periodicity of $f(t)$, the term in curly brackets in equation (\ref{P_def_osc}) oscillates with period $2\pi/\Omega$. The condition for the other oscillating term, $\exp\left[-i\left(\omega-\omega_{0}\right)t\right]$ to remain in phase with this term is
\begin{align}
\omega - \omega_{0} \approx & n\Omega \hspace{5mm}\;\;\textrm{with}\;\; n=0,\pm1,\pm2,\ldots
\label{interference_condition}
\end{align}
Reservoir modes with frequencies which do not satisfy this condition do not
contribute significantly to $\Gamma_{\infty}$ and therefore we should expect
that they do not become occupied during the atomic decay process. This is
indeed found to be the case, as outlined in subsection
\ref{reservoir_occupation_osc}.
With the above interference argument in mind and anticipating that only those
frequencies satisfying (\ref{interference_condition}) will contribute
significantly to $\Gamma_{\infty}$, we choose to re-write equation
(\ref{rewrite_Gamma_varying_gs}) as follows:
\begin{align}
\Gamma_{\infty} \propto & \lim_{t\rightarrow\infty}\Bigg\{\frac{D^{2}\gamma}{\pi{}t}\sum_{n=-\infty}^{\infty}\left\vert\int_{0}^{t}P(\omega_{0}+n\Omega,t_{1})\;dt_{1}\right\vert^{2}\Bigg\}.
\label{rewrite_Gamma_varying_gs2}
\end{align}
The proportionality constant in (\ref{rewrite_Gamma_varying_gs2}) is independent of the index $n$. In order to fix the normalisation of $\Gamma_{\infty}$, we note that the function $P(\omega_{0}+n\Omega,t)$ oscillates with period $2\pi/\Omega$, and so we may replace the long-time average by the
average over a single period:
\begin{align}
\Gamma_{\infty} \propto &
\frac{D^{2}\Omega\gamma}{2\pi^{2}}\sum_{n=-\infty}^{\infty}\left\vert\int_{0}^{\frac{2\pi}{\Omega}}P(\omega_{0}+n\Omega,t_{1})\;dt_{1}\right\vert^{2}.
\label{Gamma_nearly_varying_gs}
\end{align}
Finally, we take the dual limit:
\begin{align}
\Omega\rightarrow\infty\;\;\textrm{and}\;\;d\rightarrow0,
\label{Lambda_conditions}
\end{align}
which corresponds to a static reservoir, and must therefore return the corresponding (normalised) static decay-rate, $\Gamma_{\infty}=2D^{2}/\gamma$. This gives:
\begin{align}
\Gamma_{\infty} \approx & \sum_{n=-\infty}^{\infty}\Gamma_{n},
\label{Gamma_result_osc}
\end{align}
with the following three definitions:
\begin{align}
\Gamma_{n} \equiv 
\frac{\Omega^{2}}{2\pi}&\left\vert\int_{0}^{\frac{2\pi}{\Omega}}g^{(n)}(t)\exp\left(-i\int_{0}^{t}\left[\omega^{(n)}(\tau)-\omega_{0}\right]\;d\tau\right)\;dt\right\vert^{2}
\label{Gamma_n_osc}
 \\
&\omega^{(n)}(t) =  \omega_{0} + n\Omega + f(t)
\label{sideband_frequencies}
\\ 
&g^{(n)}(t) = \sqrt{\frac{D^{2}\gamma/\pi}{\gamma^{2}+(\omega^{(n)}(t)-\omega_{0})^{2}}}.
\label{sideband_couplings}
\end{align}
It is worth mentioning that
although we chose to move to a continuum picture from the modes, we have now arrived back at a discrete sum over reservoir variables. However, now the index ($n$) runs over reservoir sidebands (which are collectively defined), rather than individual modes ($k$) with which we began.

It is clear from equations (\ref{Gamma_result_osc}) and (\ref{Gamma_n_osc}) that in order to reduce the total decay-rate, the individual decay-channels must be inhibited, and that this can be achieved for the $n^{th}$ channel by making sure the following two conditions are satisfied:
\begin{enumerate}
\item
$g^{(n)}(t)$ is small while $\omega^{(n)}(t)$ is slowly-varying,
\item
$\omega^{(n)}(t)$ varies rapidly while $g^{(n)}(t)$ is large.
\end{enumerate}
This means that the atomic decay-rate can be inhibited if we choose the modulation function in such a way that $\vert{}f(t)/\Omega\vert$ is far from all integer-values at turning-points of $f(t)$ (see figure \ref{detuned_moving_baths}), and also $\Omega\gg\gamma$. 
%
\subsection{Reservoir occupation-spectrum}
\label{reservoir_occupation_osc}
%
%
In order to better understand the mechanism by which the atomic decay rate takes takes the form in equation (\ref{Gamma_result_osc}), it is useful to consider the final occupation spectrum of the reservoir modes. We define this spectrum
by considering those modes $k_\delta$  ($k_\delta \in \{k\}$) which
have frequencies that lie within $\delta\omega$ of an initial frequency, $\omega$. That is, we let  \cite{linington2006,cresser1983}
\begin{equation}
S(\omega) = \lim_{t\rightarrow\infty}\left\{\frac{1}{\delta\omega}\sum_{k_{\delta}}\vert{}c_{k}(t)\vert^{2}\right\}.
\label{eq:spectrum.approx}
\end{equation}
In the continuum limit the number of modes $k_\delta$ in the
sum is approximately $\delta\omega \rho(\omega)$, and since
the $c_{k}$ are expected to vary smoothly with $k$ in this
limit, we finally let
\begin{equation}
   S(\omega)  \longrightarrow  \lim_{t\rightarrow\infty}\Big\{\rho(\omega)
   |c_{k_\delta}(t)|^2\Big\},
\label{eq:spectrum.def}
\end{equation}
which applies to a representative $k_\delta$.
Equation (\ref{eq:spectrum.def}) will
serve as an operational definition of the bath spectrum.
From equation (\ref{chirp_dynamics2_osc}) we find the corresponding solution
for $c_{k}(t)$, in integral form:
\begin{widetext}
\begin{align}
\lim_{t\rightarrow\infty}\Big\{c_{k}(t)\Big\} = &
-i\int_{0}^{\infty}g_{k}(t)\exp\left[i\int_{0}^{t}\left(\omega_{k}(\tau)
  - \omega_{0} + \frac{i\Gamma_{\infty}}{2}\right)d\tau\right]\;dt,
\end{align}
which, together with the definitions (\ref{frequency_definition_osc}) and
(\ref{rho_g(t)_g(t')_osc}) and the Fourier decomposition (\ref{Fourier_decomposition}) gives
\begin{align}
S(\omega) = &
\frac{D^{2}\gamma}{\pi}\left\vert\int_{0}^{\infty}dt\sum_{m=-\infty}^{\infty}F_{m}\frac{\exp\left[i\left(\omega-\omega_{0}-m\Omega+\frac{i\Gamma_{\infty}}{2}\right)t\right]}{\sqrt{\gamma^{2}+(\omega-\omega_{0}+f(t))^{2}}}\right\vert^{2}.
\label{S_exact_osc}
\end{align}
\end{widetext}
The interference arguments used in deriving an expression for $\Gamma_{\infty}$ can be applied again
here, and so the final occupation spectrum can be written as a discrete collection of peaks at frequencies:
\begin{align}
\omega - \omega_{0} \approx & n\Omega \hspace{5mm}\;\;\textrm{with}\;\; n=0,\pm1,\pm2,\ldots
\label{interference_condition_osc2}.
\end{align}

\begin{figure}
\includegraphics[width=0.8\columnwidth]{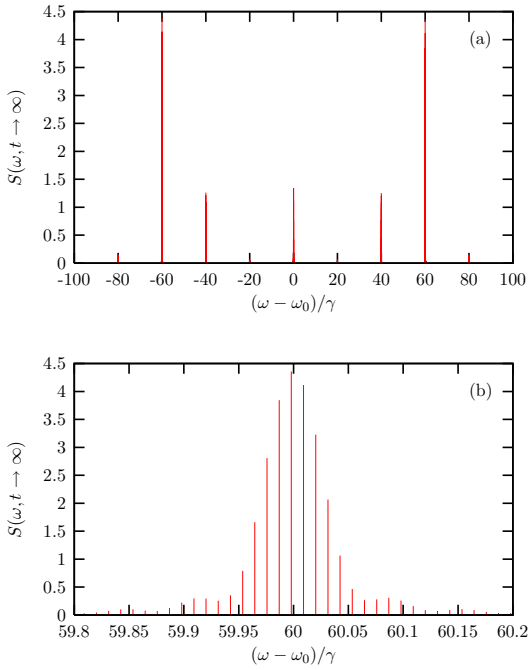}
\caption{Final reservoir occupation spectrum for the decay process studied in
section \ref{atomic_decay_rates}. The vertical axis shows the discretised form
of $S(\omega)$,
i.e. $\lim_{t\rightarrow\infty}\left\{\rho_{k}\vert{}c_{k}(t)\vert^{2}\right\}$, calculated via a
numerical solution to equations (\ref{chirp_dynamics1_osc}) and
(\ref{chirp_dynamics2_osc}). The example shown uses $D=\gamma,
\Omega=20\gamma, d=68\gamma$. (a) The final occupation spectrum
is a discrete collection of peaks, centred at $\omega_{0}+n\Omega$. This is due
to the dynamical resonance effects discussed in section \ref{dynamic_res_model}. The area
bounded by a peak indexed by $n$ is $\Gamma_{n}/\Gamma_{\infty}$. (b) The shape of each peak is approximately Lorentzian
with a half-width $\Gamma_{\infty}/2$, as predicted in section \ref{reservoir_occupation_osc}.}
\label{microbath_osc}
\end{figure}

Figure \ref{microbath_osc}, shows the final occupation spectrum, plotted using results from a numerical simulation of equations (\ref{chirp_dynamics1_osc}) and (\ref{chirp_dynamics2_osc}) for the specific case of a sinusoidal modulation (studied further in section \ref{sinusoidal_example}), which shows that the final spectrum is indeed a discrete collection of Lorentzian peaks, as predicted. The effect of dynamically manipulating the individual frequencies and coupling constants is to create a sideband structure for the reservoir. Each sideband, indexed by the integer $n$, is centred at the frequency given by equation (\ref{interference_condition_osc2}).

We write the weight of the $n^{th}$ peak as
\begin{align}
S_{n} = & \int_{\omega_{0}+(n-1/2)\Omega}^{\omega_{0}+(n+1/2)\Omega}S(\omega)\;d\omega,
\label{S_n_defn}
\end{align}
and (largely following the arguments in section \ref{atomic_decay_rates}) it is reasonably straightforward to show that
\begin{align}
S_{n} \approx & \frac{\Omega^{2}}{2\pi\Gamma_{\infty}}\left\vert\int_{0}^{\frac{2\pi}{\Omega}}g^{(n)}(t)\exp\left(-i\int_{0}^{t}\left[\omega^{(n)}(\tau)-\omega_{0}\right]\;d\tau\right)\;dt\right\vert^{2}
\\
\approx & \frac{\Gamma_{n}}{\Gamma_{\infty}}.
\label{S_n_analytic}
\end{align}
%
\section{Example: Sinusoidal modulation of reservoir mode frequencies}
\label{sinusoidal_example}
%
\subsection{Decay with sinusoidal oscillation}
\begin{figure}
\includegraphics[width=0.8\columnwidth]{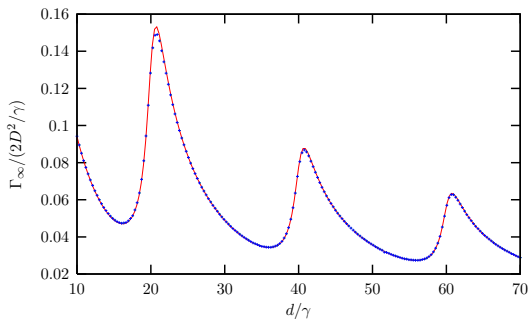}
\caption{The result of equation (\ref{Gamma_result_osc}) is plotted as a red
line and compared to numerically extracted decay rates, shown as blue
crosses. The numerically extracted decay-rates were calculated by solving
equations (\ref{chirp_dynamics1_osc}) and (\ref{chirp_dynamics2_osc}) and
fitting an exponentially decaying curve to the resulting
$\vert{}c_{a}(t)\vert^{2}$. The modulation frequency $\Omega=20\gamma$.}
\label{Gamma_osc_fig}
\end{figure}

%
The analysis of the previous section is fairly general, in that the sole requirement placed on the reservoir manipulation is that the modulation function $f(t)$ is periodic in time. In order to present a concrete test of the predicted decay-rates and reservoir occupation spectrum in the current section, we apply our results to the simplest periodic case - namely a sinusoidal frequency manipulation, with modulation depth $d$ and modulation frequency $\Omega$
\begin{align}
f(t) = & d\sin(\Omega{}t).
\end{align}
For this case, the Fourier decomposition of the phase-term is \cite{abramowitz1972}:
\begin{align}
\exp\left[id\left(\cos(\Omega{}t)-1\right)/\Omega\right]
= & e^{-id/\Omega}\sum_{n=-\infty}^{\infty}i^{n}J_{n}(d/\Omega)e^{in\Omega{}t},
\end{align}
where $J_{n}$ is the $n^{th}$-order Bessel function of the first kind.
The decay-rate into the $n^{th}$ sideband given in equation (\ref{Gamma_n_osc}) thus takes the form:
\begin{widetext}
\begin{align}
\Gamma_{n} = &
\frac{D^{2}\Omega^{2}\gamma}{2\pi^{2}}\left\vert\sum_{m=-\infty}^{\infty}i^{m}J_{m}(d/\Omega)\int_{0}^{\frac{2\pi}{\Omega}}\frac{e^{-i\left(n
    -
    m\right)\Omega{}t}}{\sqrt{\gamma^2+\left(n\Omega+d\sin(\Omega{}t)\right)^{2}}}\;dt\right\vert^{2}.
\label{Gamma_n_osc_ex}
\end{align}
\end{widetext}
The total atomic decay-rate, $\Gamma_{\infty}=\sum_{n}\Gamma_{n}$ is plotted in figure \ref{Gamma_osc_fig} and compared with a direct numerical solution of equations (\ref{chirp_dynamics1_osc}) and (\ref{chirp_dynamics2_osc}) for a range of modulation depths and fixed modulation frequency. As explained in the text of section \ref{atomic_decay_rates}, $\Gamma_{\infty}$ is a peaked function of the modulation depth, with a maximum every time $d$ is an integer-multiple of $\Omega$. 
%
%

The final reservoir occupation spectrum, $S(\omega)$, calculated numerically from a direct simulation of equations (\ref{chirp_dynamics1_osc}) and (\ref{chirp_dynamics2_osc}) is plotted in figure \ref{microbath_osc}. As predicted, this spectrum is a collection of Lorentzian peaks centred at angular frequencies $\omega\approx\omega_{0}+n\Omega$.

\begin{figure}[]
\includegraphics[width=0.9\columnwidth]{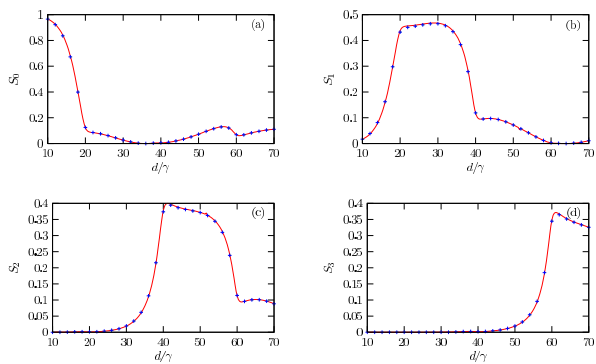} 
\caption{Final occupation peak-weights, $S_{n}$ as a
function of the modulation-depth, $d$ for the first four peaks. The figures
\mbox{(a)-(d)} correspond to \mbox{$n=0,1,2,3$} respectively. The $n^{th}$
peak weight is predicted to be $\Gamma_{n}/\Gamma_{\infty}$, with $\Gamma_{n}$
given by (\ref{Gamma_n_osc}). As discussed in subsection \ref{atomic_decay_rates},
$S_{n}\sim0$ for $d<n\Omega$ and
$S_{n}$ peaks at $d\sim{}n\Omega$.}
\label{S_n_peaks} 
\end{figure}

In order to test the predictions for the weight, $S_{n}=\Gamma_{n}/\Gamma_{\infty}$, of each peak, figure \ref{S_n_peaks} shows the analytic result of equation (\ref{S_n_analytic}) together with the numerically calculated peak weights defined in equation (\ref{S_n_defn}), for the same model parameters as figure \ref{Gamma_osc_fig}.

\subsection{Decay in the limit of ultra-fast reservoir modulation}
\label{ultrafast}
%
The simplest analytic expression for $\Gamma_{\infty}$ is obtained in the limit
that the modulation frequency dominates over all other frequency-scales in the
system (except for the free-evolution):
\begin{align}
\omega_{0} \gg \Omega \gg d, \gamma, D.
\label{high_Omega_osc}
\end{align}
In this limit, the denominator of every term is small for $n\neq0$. Also, the sum over $m$ is dominated by the $m=0$ term, due to the factor $J_{m}(d/\Omega)$, and so we only need to consider the $n=m=0$ term. This can be tackled analytically:
\begin{align}
\Gamma_{\infty} \approx &
\frac{D^{2}\Omega^{2}\gamma}{2\pi^{2}}\left\vert{}J_{0}(d/\Omega)\int_{0}^{\frac{2\pi}{\Omega}}\frac{1}{\sqrt{\gamma^{2}+d^{2}\sin^{2}(\Omega{}t)}}\;dt\right\vert^{2}
\nonumber \\
\approx & 
\frac{2D^{2}}{\gamma}\left\{\frac{\left[2\gamma{}J_{0}(d/\Omega)\mathcal{K}\left(\frac{d^{2}}{\gamma^{2}+d^{2}}\right)\right]^{2}}{\pi\left(\gamma^{2}+d^{2}\right)}\right\},
\label{analytic_ours}
\end{align}
where $\mathcal{K}$ represents the complete elliptic integral of the first
kind \cite{abramowitz1972}. In this limit, $\Gamma_{\infty}$ is a monotonically decreasing function of $d/\gamma$ and the inhibition of atomic decay can be quite dramatic, as illustrated in the following section.
%
\section{Physical Realisation and Observable Effects}
\label{observable_effects}
%
\subsection{Double-cavity system}
\label{double_cavity_system}
%
The reservoir manipulation studied in this paper is quite specific in the sense that the microscopic reservoir modes have time-dependent frequencies and couplings to the atom, but together, they must conspire to keep the macroscopic reservoir structure fixed in time. This precise form was deliberately chosen in order to isolate the new effects arising here from other previously studied types of engineered reservoir, such as \cite{kofman2001, myatt2000, turchette2000, celeri2006}

\begin{figure}[!htb]
\begin{center}
\includegraphics[width=0.7\columnwidth]{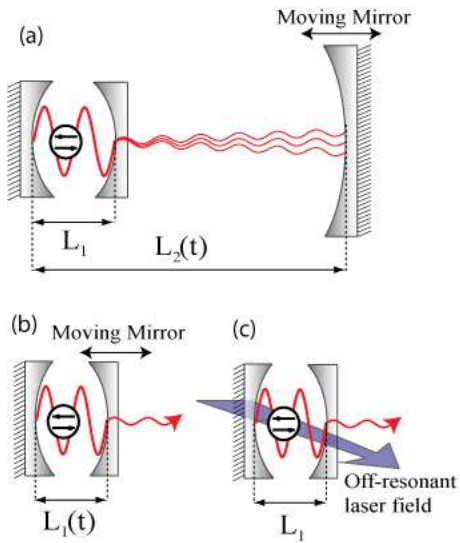}
\caption{(a) To manipulate the microscopic mode-structure of the reservoir
(double-cavity), the outer-cavity mirrors are moved, which affects the mode
frequencies as described by (\ref{frequency_definition_osc}).
Since the inner-cavity mirrors
are fixed, the macroscopic properties of the reservoir remain invariant which
recovers the Hamiltonian of equation (\ref{Hamiltonian}).
(b) Modulating the length of the cavity itself acts to dynamically detune the cavity resonance from the atomic transition, resulting in the interaction Hamiltonian (\ref{H_int_osc_Janowicz}). (c) Alternatively, variable detuning between the atom and reservoir may be achieved by inducing a time-dependent Stark-shift in the atomic transition frequency \cite{noel1998}. This method also results in the Hamiltonian given in (\ref{H_int_osc_Janowicz}).}
\label{double_cavity_figure}
\end{center}
\end{figure}

In a recent paper, we suggested one way of physically implementing such a dynamic environment \cite{linington2006}; the atom is placed inside
a static cavity, which itself is enclosed in a larger cavity with moving boundaries, as in figure \ref{double_cavity_figure}(a). In this way, the envelope of the reservoir structure is determined by the properties of the inner cavity, and so remains fixed. On the other hand, the frequencies of individual modes are fixed by satisfying a node condition on the \emph{outer} mirrors, and so these can be modulated by moving the right-hand mirror.

%
\subsection{Comparison with a variable-detuning model}
\label{variable_detuning}
%
It was noted in section \ref{sinusoidal_example} (e.g. figure \ref{Gamma_osc_fig}) that dynamically manipulating the reservoir structure in the way described here can result in an inhibition of the atomic decay rate, despite the fact that the atom and reservoir are weakly coupled, and always remain on resonance during the decay. In this subsection, we choose to highlight that the extent of this inhibition is far greater than might first be expected. 
To this end, we choose to contrast the situation studied in sections \ref{general_dynamic_environment}-\ref{sinusoidal_example} with  with the case of a \emph{variable atom-reservoir detuning}, i.e.
\begin{align}
\hat{H}_{I}'(t) = &
\sum_{k}g_{k}\bigg[\hat{\sigma}^{-}\hat{b}_{k}^{\dagger}\exp\left(i\int_{0}^{t}
\big[\omega_k(\tau) - \omega_{0}\big]
d\tau\right)  
\nonumber \\
&  +\hat{\sigma}^{+}\hat{b}_{k}\exp\left(-i\int_{0}^{t}\big[\omega_k(\tau') - \omega_{0}\big]d\tau'\right)\bigg].
\label{H_int_osc_Janowicz}
\end{align}
Modulating either one of the atomic transition frequency (figure \ref{double_cavity_figure}(c)) or the central frequency of the reservoir (figure \ref{double_cavity_figure}(b)) would give rise to the interaction Hamiltonian in equation (\ref{H_int_osc_Janowicz}). 
In either case, the coupling strengths $g_{k}$ are static and do not compensate for the changing mode-frequencies and this has the effect of dynamically modulating the macroscopic atom-reservoir detuning. Such variable-detuning models have been well studied (see for example \cite{janowicz2000, agarwal1999, law1995, garraway1997b}) and in the weak-coupling limit the atomic decay-rate is given by \cite{agarwal1999}:
\begin{align}
\Gamma_{\infty}' \approx & \sum_{n}\Gamma_{n}'
\\
\Gamma_{n}' = &
\frac{2D^{2}}{\gamma}J_{n}(d/\Omega)^{2}\left(\frac{\gamma^{2}}{\gamma^{2}+n^{2}\Omega^{2}}\right).
    \label{Gamma_Janowicz}
\end{align}
In the ultrafast modulation limit considered in section \ref{ultrafast}, equation (\ref{Gamma_Janowicz}) simplifies to
\begin{align}
\Gamma_{\infty}' \approx & \frac{2D^{2}}{\gamma}J_{0}(d/\Omega)^{2}.    
\label{analytic_Janowicz}
\end{align}
Somewhat surprisingly, in the limit of ultrafast frequency modulation, the decay-rate $\Gamma_{\infty}'$ is \emph{always greater than} the equivalent decay rate $\Gamma_{\infty}$ given in equation (\ref{analytic_ours}) for the same frequency modulation and depth. Therefore a more effective suppression of the atomic decay-rate can be achieved by keeping the atom and reservoir on-resonance rather than dynamically changing the detuning.

\begin{figure} [h!]
\includegraphics[width=0.8\columnwidth]{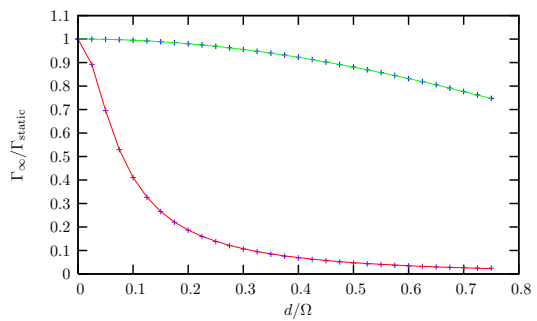}
\caption{Analytic results,  (\ref{Gamma_n_osc_ex}) and (\ref{Gamma_Janowicz}) (crosses), together with the leading-order terms, (\ref{analytic_ours}) and (\ref{analytic_Janowicz}) (solid lines). The modulation frequency $\Omega=20\gamma$.
In this limit, the reservoir manipulation studied in our paper (lower curve) is more efficient at suppressing dissipative effects than an oscillatory detuning (upper curve).}
\label{analytic_comparison}
\end{figure}

The approximate analytic results given in equation (\ref{analytic_ours}) and equation (\ref{analytic_Janowicz}) are compared in figure \ref{analytic_comparison} together with the full decay-rates given in equations (\ref{analytic_ours}) and (\ref{Gamma_Janowicz}). 
%
%

We note also that the ratio between the decay rate derived in this paper (section \ref{ultrafast}) and the decay rate arising in a variable-detuning model may be written as
\begin{align}
\frac{\Gamma_{\infty}}{\Gamma_{\infty}'} \equiv f(x) = \frac{4\left[\mathcal{K}\left(\frac{x^{2}}{1+x^{2}}\right)\right]^{2}}{\pi^{2}(1+x^{2})},
\label{f(x)}
\end{align}
with $x=d/\gamma$. This is a monotonically decreasing function of $x$, and always less than 1. Providing that the modulation frequency is high enough, it is possible to achieve high values of $x$ whilst still satisfying $\Omega\gg{}d$.

%
\subsection{Size of effects}
\label{size_effects}
%
Experimental verification of the predictions made above may be achieved with currently available technology and without great difficulties. First we take an optical atomic
transition with angular frequency \mbox{$\omega_{0}/2\pi\sim3.5\times10^{14}$ Hz} and a 40 $\mu$m inner
cavity for which \mbox{$\gamma/2\pi\sim4.1$ MHz} \cite{hood2001, boca2004, miller2005}, whilst for the outer cavity we let $L=$ 1cm. Assuming that the end-mirror is driven sinusoidally by a piezoelectric actuator, with frequency $\Omega/2\pi\sim40$MHz and an amplitude $d\sim{}0.4$nm, the ratio $\Omega>d>\gamma$ is achieved and the resulting decay rate, $\Gamma_{\infty}$ is four times lower than for a variable-detuning model.

For microwave systems, we consider an atomic transition with frequency \mbox{$\omega_{0}/2\pi\sim21$GHz} and a 2cm inner cavity for which \mbox{$\gamma/2\pi\sim10$ Hz} \cite{varcoe2000}, whilst for the outer cavity we let $L\sim20$cm. In this case, a frequency $\Omega/2\pi\sim1$MHz and an amplitude $d\sim{}10$nm, gives \mbox{$\Omega/d=d/\gamma=1000$} so the heirachy given in equation (\ref{high_Omega_osc}) can be made to hold very strongly.
The resulting inhibition of the decay-rate (equation (\ref{f(x)})) is \mbox{$\Gamma_{\infty}\sim2.8\times10^{-5}\Gamma_{\infty}'$}; decay into the cavity field mode is effectively `switched-off' in this limit \footnote{In this regime, a thorough assessment of the atomic decay-rates would require us to include other, lower-order effects into the model, such as spontaneous emission into other background modes.}.

\section{Conclusions}
\label{conclusions}
%
We have studied  how the process of spontaneous emission may be controlled by dynamically manipulating the microscopic structure of an atom's environment. While is it well known that changing the structure of an atom's surroundings (by introducing nearby conducting surfaces or a cavity, for example) can alter the process of spontaneous emission, in this article we have considered a model for which the macroscopic reservoir-structure \emph{does not} change with time. Instead, the individual electromagnetic field-modes which make up the reservoir are assigned time-dependent frequencies and time-dependent couplings to the atom. In this way, the atom and reservoir always remain on resonance but the atom interacts with different bath-modes at different times; by manipulating the environment in this manner it is possible to control the memory-kernel for the atomic state and hence to control the atomic decay-rate.

In order for such a scheme to work, we must be able to alter the reservoir-structure on a timescale which is shorter than the static-bath correlation time. Therefore, it is essential that the system-bath couplings are frequency-dependent, as shown in section \ref{general_dynamic_environment}. Although essential, this frequency-dependence may be very weak, and in section \ref{dynamic_res_model} we show that it is possible to control the atomic decay-rate even in the case of weak atom-reservoir coupling (as well as in the strong-coupling limit -- see \cite{linington2006}).
In subsection \ref{atomic_decay_rates} the atomic decay-rate is calculated for the case of a periodic manipulation of all reservoir mode-frequencies. A Floquet analysis is applied and after analysing the final occupation-spectrum (in subsection \ref{reservoir_occupation_osc}), we find that the atomic decay only occurs into a discrete collection of reservoir-sidebands, each of which satisfies a dynamic resonance condition with the atom. The total atomic decay-rate is a structured function of the modulation-depth and modulation-frequency and exhibits a peak whenever one of the reservoir sidebands is resonant with the atom for a significant fraction of each modulation cycle. 

Section \ref{sinusoidal_example} treats the specific case of a sinusoidal modulation of all reservoir mode-frequencies and the general expressions derived in section \ref{dynamic_res_model} are found to be in excellent agreement with numerical simulations. Of particular interest is the limit of ultrafast modulation, studied in subsection \ref{ultrafast}, for which compact analytic expressions are derived.

In section \ref{observable_effects} the observable effects of the model are considered. In order to emphasise how strong the suppression of decay may be in the  limit of ultrafast modulation, a second model is introduced in subsection \ref{variable_detuning} for which the atom-reservoir detuning is modulated sinusoidally. Upon first inspection, it may appear that this second reservoir manipulation should give rise to a more dramatic inhibition of the atomic decay, since in this case the atom and reservoir spend a large proportion of each cycle away from resonance. However, a detailed analysis shows that the dynamical suppression of decay studied in sections \ref{dynamic_res_model} and \ref{sinusoidal_example} is a far more potent method for inhibiting dissipation than a straightforward atom-bath detuning.
Physical realisations for both of these types of dynamic environment are proposed in subsection \ref{double_cavity_system} (also, more details are given in \cite{linington2006}). Finally, in subsection \ref{size_effects}, we show that it should be possible to observe all of the effects predicted above using currently available technology.

\begin{acknowledgments}
 We would like to thank and acknowledge support from the
   Leverhulme Trust, the UK Engineering and Physical Sciences Research Council, the EU ToK project CAMEL (Grant No. MTKD-CT-2004-014427) and the EU RTN project EMALI (Grant No. MRTN-CT-2006-035369).
\end{acknowledgments}


\end{document}